\documentclass[useAMS,usenatbib,babel]{mn2e}

\usepackage[english,english]{babel}
\usepackage{amsmath}
\usepackage{amssymb,amsfonts,textcomp}
\usepackage{array}
\usepackage{supertabular}
\usepackage{hhline}
\usepackage{hyperref}
\usepackage[usenames]{color}

\usepackage{textcomp}

\usepackage{graphicx}

\usepackage{dcolumn}
\usepackage{bm}
\usepackage{comment}
\usepackage{color}

\newcommand{\mP}{{\cal P}}

\newcommand{\dd}{\hbox{d}}

\DeclareMathOperator*{\argmax}{argmax}

\usepackage{color}
\definecolor{Red}{rgb}{0.65,0.08,0.05}
\def\red{}

\begin{document}
\title[Constraining dark energy via density in spheres]{Encircling the dark:\\ constraining dark energy via cosmic density in spheres
 }
\author[S.~Codis, C.~Pichon, F.~Bernardeau, C.~Uhlemann and S.~Prunet
]{
\parbox[t]{\textwidth}{
 S.~Codis$^{1}$,  C.~Pichon$^{2,3}$, F.~Bernardeau$^{2,4}$, C.~Uhlemann$^{5}$ and S.~Prunet$^{6}$
}
\vspace*{6pt}\\
\noindent
$^{1}$ Canadian Institute for Theoretical Astrophysics, University of Toronto, 60 St. George Street, Toronto, ON M5S 3H8, Canada\\
$^{2}$ Sorbonne Universit\'es, UPMC Paris 6 \& CNRS, UMR 7095, Institut d'Astrophysique de Paris,
 98 bis bd Arago, 75014 Paris, France\\
 $^{3}$ Korea Institute of Advanced Studies (KIAS) 85 Hoegiro, Dongdaemun-gu, Seoul, 02455, Republic of Korea\\
$^{4}$ CNRS \& CEA, UMR 3681, Institut de Physique Th\'eorique, F-91191 Gif-sur-Yvette, France\\
$^{5}$  {Institute for Theoretical Physics, Utrecht University, Leuvenlaan 4, 3584 CE Utrecht, The Netherlands}\\
$^{6}$  Canada France Hawaii Telescope Corporation  65-1238 Mamalahoa Hwy  Kamuela, Hawaii 96743  USA
}

\maketitle
\begin{abstract}
The recently published  analytic  probability density function for the mildly non-linear cosmic density field within
 spherical cells is used to build a simple but
 accurate maximum likelihood estimate for the redshift  evolution of the variance of the density, which, as expected, is shown to have smaller relative error
 than the sample variance.
This estimator provides a competitive probe for the equation of state of dark energy, reaching a few percent  accuracy on $w_{p}$ and $w_a$ for a Euclid-like survey. 
 The corresponding likelihood function can take into account the configuration  of the cells via their relative separations.
A code to compute one-cell density probability density functions for arbitrary initial power spectrum, top-hat smoothing and various spherical collapse dynamics is made available online so as to provide straightforward means of testing the effect of alternative dark energy models and 
initial power-spectra on the low-redshift matter distribution.
\end{abstract}

\section{Introduction}
{Soon after the observation of accelerated expansion of the Universe in supernova data in \cite{Riess98}, along with measurements of the CMB anisotropy, large-scale structure (LSS) provided independent evidence for dark energy and allowed for improved constraints \citep[see][]{Perlmutter99}. Since then several well-established and complementary methods are pursued to accurately measure expansion history and structure growth \citep[see][ for a review]{Weinberg13}.}
With the advent of large galaxy surveys covering a significant fraction of the sky (e.g. SDSS and in the coming years Euclid~\citep{Euclid}, DES \citep{Abbott:2005bi}, eBOSS \citep{eBOSS}, and LSST \citep{lsst}), astronomers are  attempting to understand the origin of dark energy. One venue is to  build tools that can  extract  accurate information  from these data sets as a function of redshift during the epoch where dark energy dominates. {Constraints can be obtained from direct observations of distance measures or the growth of structure which is not only sensitive to dark energy but also modified gravity. Baryon acoustic oscillations (BAOs) provide a standard ruler imprinted in the galaxy correlation function such that measuring the preferred clustering scales in redshift and angular separations allows to determine the angular diameter distance to and Hubble parameter at the redshifts of the survey, respectively. } 
{Measuring the growth of structure demands probing the non-linear regime of structure formation, for example by means of galaxy correlation functions. } 
{Recently, topological and geometrical estimators restricted to some special locations of the cosmic web have attracted attention as a potential dark energy probe accessible from galaxy surveys, for example the geometry of the filaments \citep{Gay2012} or Minkowski functionals \citep[e.g][for a recent implementation in redshift space]{Codis2013} to name a few.
Here, we will present the probability distribution function (PDF) of the density in spherical cells as a complementary probe to constrain dark energy.
The idea to probe subsets of the matter field, such as over- and under-densities, is an interesting direction that can also be investigated in a count-in-cell formalism.}

In this letter, we  illustrate how recent progress in the context of large deviations \citep[e.g.][]{LDPinLSS} allow for 
the construction of estimators for the equation of state of dark energy relying on an
explicit analytic expression for the PDF of the density in spherical cells as a function of redshift \citep{Uhlemann16}. 
The emphasis is on simplicity and proof of concept rather than realism. 
More precisely, we give a simple but very accurate form of the density PDF in terms of the amplitude $\sigma(R,z)$ of the density fluctuation at radius $R$. We then relate the change of  variance $\sigma(R,z)$ with redshift $z$ to the linear growth rate $D(z)$ which is parametrized by the dark energy parameters $(w_a,w_{p})$ that characterize the equation of state $w(z) = w_{p} + w_a/(1 + z)$\citep{Euclid}. A maximum likelihood estimator is built from the analytical density PDF to extract a constraint on $(w_a,w_{p})$. 
 We present and distribute the corresponding code for wider use {and application to different linear power spectra,
 using e.g. {\tt camb} \citep{camb}. Furthermore, we give an outlook of how the framework allows to account for the extent of the survey and model the 
corresponding biases.} 

This paper is organized as follows. Section~\ref{sec:PDF}  
presents  the analytical PDF of the density in spherical cells {parametrized in terms of the underlying variance}.
Section~\ref{sec:sample} compares the sample versus the maximum likelihood variance.
Section~\ref{sec:fiducial} {describes a fiducial dark energy experiment for an Euclid-like survey and presents the resulting constraints on the equation of state parameters}.
Section~\ref{sec:conclusion} wraps up.
Importantly, Appendix~\ref{sec:code} provides links to the relevant codes 
 for arbitrary power-spectra, while Appendix~\ref{sec:comparing}  compares the two different estimators of the variance of the density field
 using the asymptotic Fisher information.

\begin{figure}
\includegraphics[width=\columnwidth]{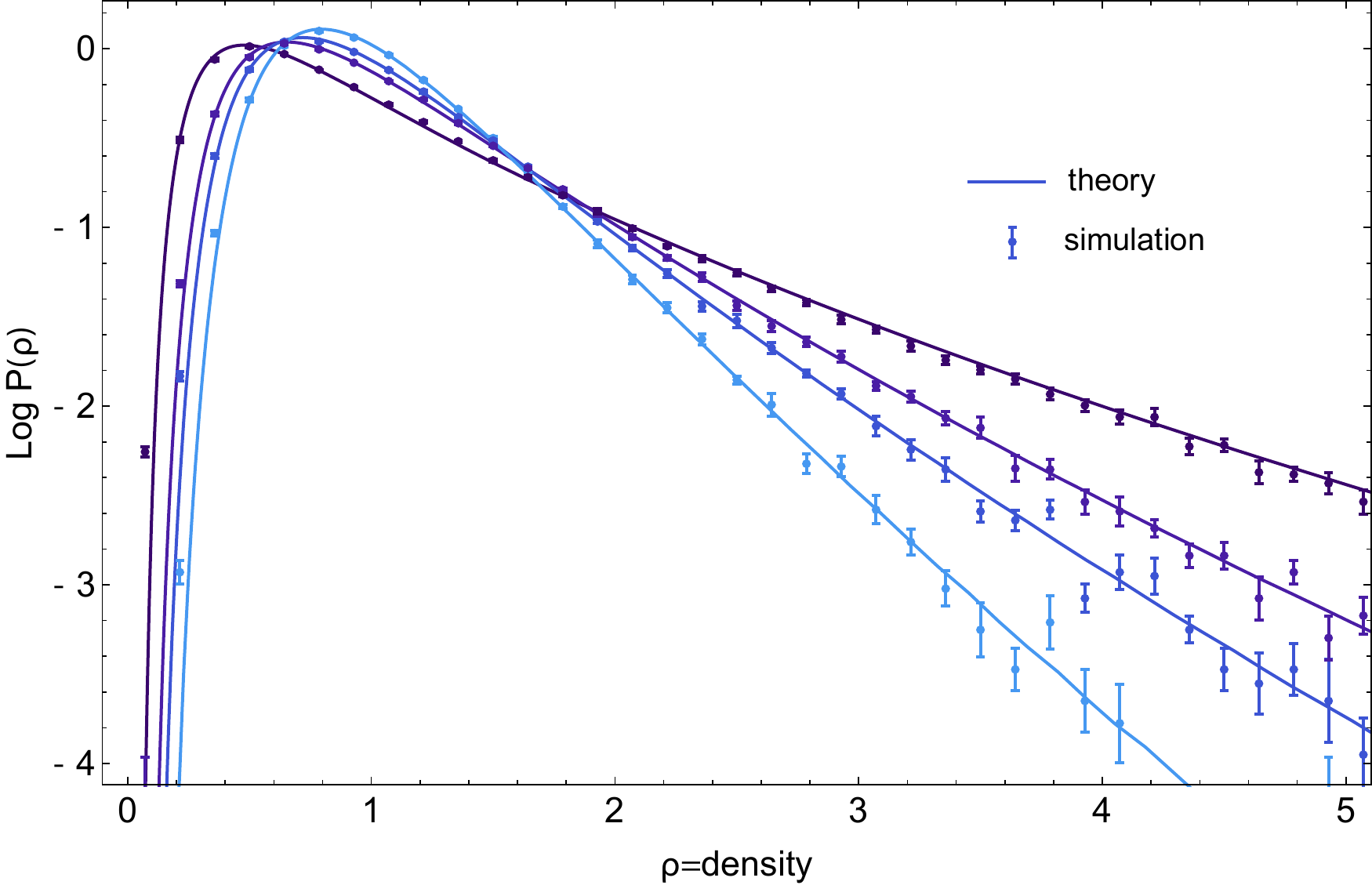}
   \caption{
   PDF of $\rho$ measured (error bars) and predicted via 
    a saddle point approximation in the PDF of $\log \rho$ given by equation~\eqref{PDFfromPsi2}. Four different redshifts are shown : $z=1.36$, 0.97, 0.65 and 0 (from light to dark blue), for a filtering scale $R=10$ Mpc/$h$.
   \label{fig:fit-PDF}}
\end{figure}

\section{The analytic density PDF}
\label{sec:PDF}

It has been shown in \cite{Uhlemann16} that the PDF, $\mP_{R}(\rho|\sigma) $, for the density within a sphere of radius $R$ at redshift $z$ (with corresponding variance $\sigma\equiv\sigma(R,z)$) has a simple analytical expression given 
by
\begin{equation}
\label{PDFfromPsi2}
\mP_{R}(\rho|\sigma) = \sqrt{\frac{\Psi''_R(\rho)+\Psi'_R(\rho)/\rho}{2\pi \sigma^{2}}} \exp\left(-\frac{\Psi_R(\rho)}{\sigma^{2}}\right)\,,
\end{equation}
where prime stands for derivative w.r.t. $\rho$ and
\begin{equation}
\label{Psiquad}
 \Psi_R(\rho)= \frac{\sigma^2}{2\sigma^2(R\rho^{1/3})} \tau(\rho)^2\,,
\end{equation}
with $\tau(\rho)$ the linear density contrast within the Lagrangian radius $ R\rho^{1/3}$. According to the spherical collapse model $\tau(\rho)=\zeta^{-1}(\rho)$ can be expressed as a non-linear transform of the density  $\rho$ within radius $R$ via an accurate  fit for $\zeta(\tau)$ given by
\begin{equation}
\rho=\zeta(\tau)=(1-\tau/\nu)^{-\nu}\,, \label{eq:spherical-collapse}
\end{equation}
presented in \cite{Bernardeau92} for $\nu=3/2$. Here $\nu=21/13$ is chosen to match the exact high redshift skewness as described in \cite{Bernardeau14}.
The density PDF given by equation~(\ref{PDFfromPsi2}) depends on redshift only through $\sigma=\sigma(R, z)$, the amplitude of the fluctuation at scale $R$ and redshift $z$. In this work, $\sigma(R,z)$ will be assumed to scale like 
the growth rate function, $D(z)$, although this strictly speaking holds only in the linear regime.
In particular, note that in equation~(\ref{Psiquad}), $\sigma(R\rho^{1/3})/\sigma$ is a function of the density $\rho$, and cell size $R$ but does not depend on  redshift 
 \begin{equation}
\frac{ \sigma^2(\rho^{1/3} R)}{\sigma^{2}}=  \frac{\displaystyle \int  P^{\rm lin}_k(k) W^2(\rho^{1/3} R k) \dd^3k}{\displaystyle\int  P^{\rm lin}_k(k) W^2(R k) \dd^3k}\,, \label{eq:defsigma}
 \end{equation}
 where  $ W(k)={3}/{k^{2}}  {\left({\sin(k)}/{k}-\cos(k)\right)}$ is the top-hat filter at scale $R$.
 Equation~(\ref{eq:defsigma}) encodes the dependency of equation~(\ref{PDFfromPsi2}) w.r.t. the initial power spectrum. 
 
The above prescription yields simple analytic expression for the PDF. 
 For instance, for a scale-invariant power spectrum of index $n$ {($<2.6$)}, and spherical collapse factor $\nu$,
 the (un-normalised) PDF has this simple  form
 \begin{multline}
\mP_{R}(\rho|\sigma)=
 \exp\left[{-\frac{\nu ^2 \left(\rho ^{\frac{1}{\nu }}-1\right)^2 \rho
   ^{-\frac{2}{\nu }+\frac{n}{3}+1}}{2 \sigma ^2}}\right]\times
   \\
   \sqrt{\!\frac{\nu ^2 (n+3)^2\! \left(\!\rho ^{\frac{1}{\nu
   }}\!-\!1\!\right)^2\!\!\!\!+\!12 \nu  (n+3)\! \left(\!\rho ^{\frac{1}{\nu }}\!-\!1\!\right)\!\!-\!18 \!\left(\!\rho ^{\frac{1}{\nu }}\!-\!2\!\right)}{36 \pi  \sigma^{2}\rho ^{\frac{2}{\nu }+1-\frac{n}{3}} }\!}.
   \end{multline} 
 Note that, as suggested by \cite{Uhlemann16}, the density PDF given by equation~(\ref{PDFfromPsi2}) must be normalised by dividing the PDF by its integral and the density by its mean.
 
 Figure~\ref{fig:fit-PDF} shows the comparison between the predicted PDF given by equation~(\ref{PDFfromPsi2}), when only the spectral index and running of the linear power spectrum at the scale of interest ($R=10$Mpc$/h$) are considered, and the density PDF measured in a N-body simulation
 \citep[see][for details]{Bernardeau14}. Percent accuracy is reached for $\sigma$ below $0.7$.
\begin{figure*}
\includegraphics[width=\columnwidth]{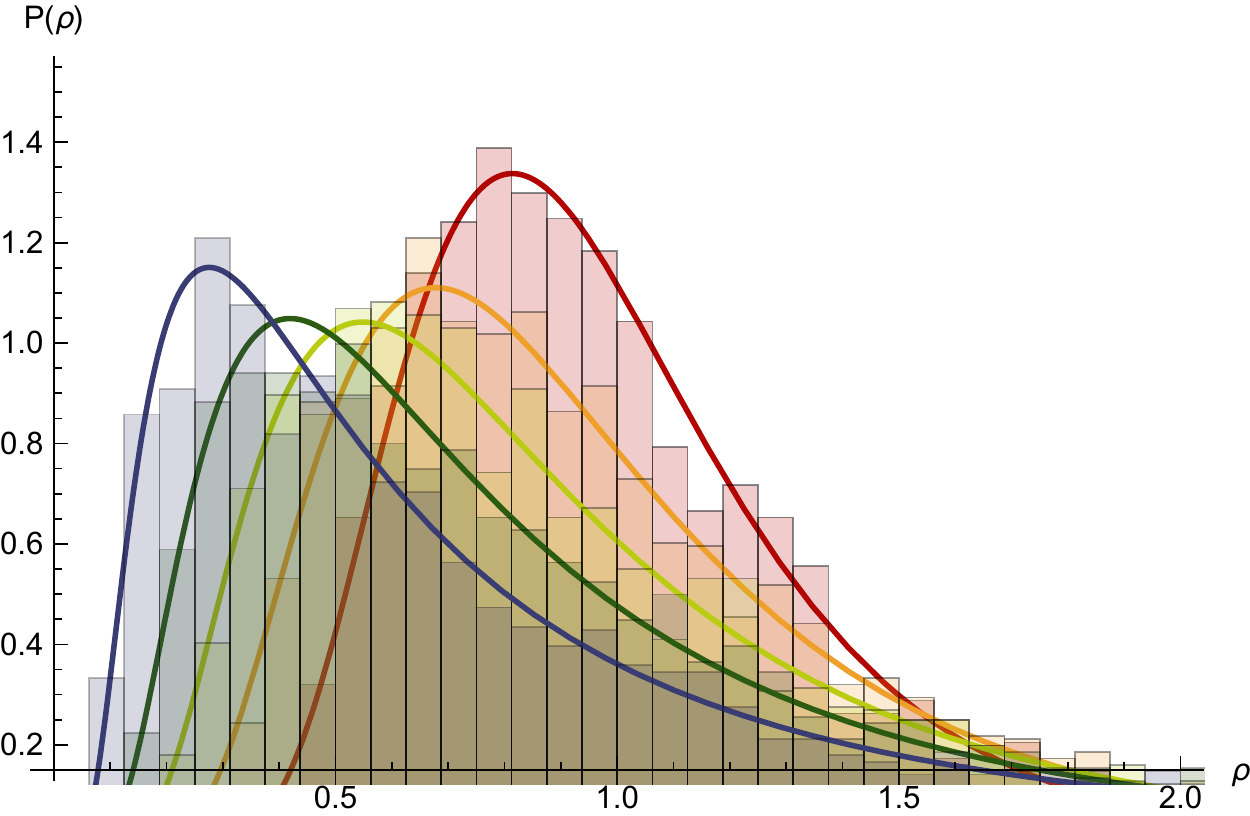}
\includegraphics[width=\columnwidth]{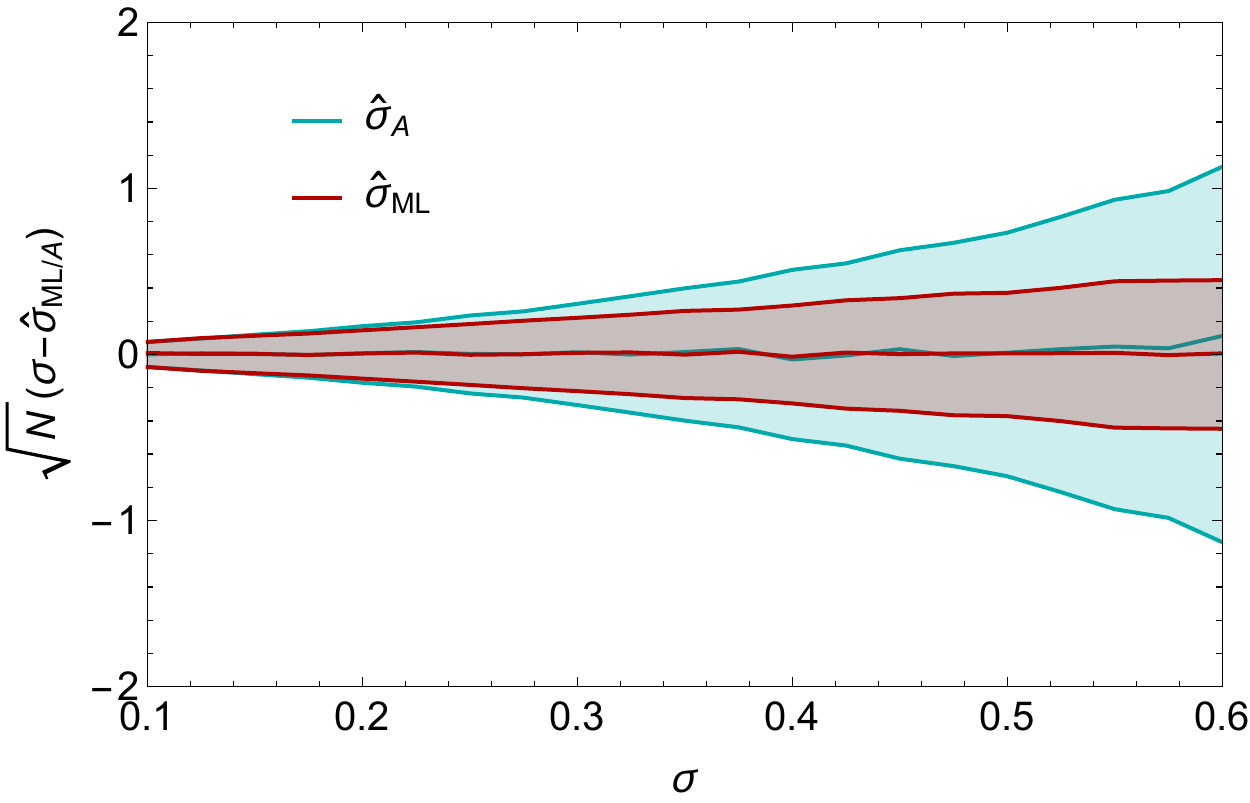}
   \caption{
   {\sl Left-hand panel}: the expected and sampled PDF for different values of $\sigma$ when $N=2500$ densities are randomly drawn for each density variance.
   {\sl Right-hand panel}: mean (and standard deviation) of the density variance estimated from the 
   arithmetic estimator ($\hat{\sigma}_{\rm A}$) in cyan compared to the variance of the maximum likelihood estimator ($\hat{\sigma}_{\rm ML}$) in red, estimated by equations~(\ref{eq:sigA}) and ~(\ref{eq:sigML}). 1000 realisations of this experiment are used to compute the mean (solid line) and standard deviation (shaded area). 
   As expected, in the zero variance limit where the PDF is close to Gaussian, both estimators are equivalent but the maximum likelihood approach  clearly  outperforms the sample variance estimate as $\sigma$ increases (by factor $2.5$ at $\sigma=0.6$).
   \label{fig:fit-error}}
\end{figure*}

 In practice, equation~(\ref{PDFfromPsi2}) can be applied to {\it arbitrary} linear power spectra. Note that this prediction for the density PDF depends on the cosmological model through i) the linear power spectrum and ii) the dynamics of the spherical collapse (parametrized by $\nu$ here, see equation~\eqref{eq:spherical-collapse}).  Appendix~\ref{sec:code} presents the code {\tt LSSFast} to compute such PDFs together with a bundle of PDFs for different variances and radii.

\section{ML versus sample variance}
\label{sec:sample}

Because we have an accurate theoretical model for the full density PDF given by equation~(\ref{PDFfromPsi2}), it is now possible to build a maximum likelihood estimator for the density variance.
In order to compare this approach to the traditional sample variance, we
consider a set of $2,500$ spheres of radius $R=10$Mpc$/h$ for 13 different redshifts (and therefore 13 different density variance $\sigma_{k}=\sigma(z_{k},R)$),
and we draw for each sphere a random density, $\{\rho_{i,k}\}$.

Figure~\ref{fig:fit-error}, left-hand panel, displays the predicted PDFs as well as the corresponding draws.
The right-hand panel displays the corresponding relative error (mean and standard deviation) on the estimate of $\sigma$ either from 
the maximum likelihood estimator, or while computing directly the 
variance of each redshift sample based on 100 Monte Carlo realisations of this experiment.
Since the theory provides us with the expected PDF, the maximum likelihood based on this PDF is optimal and unbiased. While the sample variance is competitive at small variance when the PDF is nearly Gaussian, it becomes clearly sub-optimal for larger values of the density variance, e.g by a factor of three for $\sigma=0.7$.
Appendix~\ref{sec:comparing} shows analytical results on the comparison between sample variance and maximum likelihood estimator.
 in the asymptotic limit where $N$ goes to infinity.

\section{Fiducial dark energy experiment}
\label{sec:fiducial}

The redshift evolution of the underlying density variance $\sigma^{2}(R,z)$
can then be used to pin down the parameters of the equation of state of dark energy. 
Indeed, a dark energy probe may directly attempt to estimate the so-called equation of state parameters $(w_a,w_{p})$ from the PDF  while relying on
 the cosmic model for the growth rate \citep{Glazebrook}, 
\begin{align}
D(z|w_{p},w_a)&=\frac{5\Omega_m H_0^2}{2} H(a)\int_0^a \frac{{\rm d} a'}{a'^3 H^3(a')}\,, \\
H^2(a)=H_0^2 &\left[ \frac{\Omega_m}{a^3}+ \Omega_\Lambda \exp\left(3 \int_0^z \frac{1+w(z')}{1+z'} {\rm d} z'  \right)
 \right], \, \label{eq:cosmo}
\end{align}
 with $\Omega_m$, $\Omega_\Lambda$ and $H_0$ resp. the dark matter and dark energy  densities and the Hubble constant at redshift $z=0$,  $a\equiv1/(1+z)$ the expansion factor, and with the equation of state $w(z)=w_p+ w_a /({1+z})$. {Note that the same approach was employed by \cite{Gay2012} to obtain a dark energy constraint based on geometrical critical sets.}

\begin{figure}
\includegraphics[width=\columnwidth]{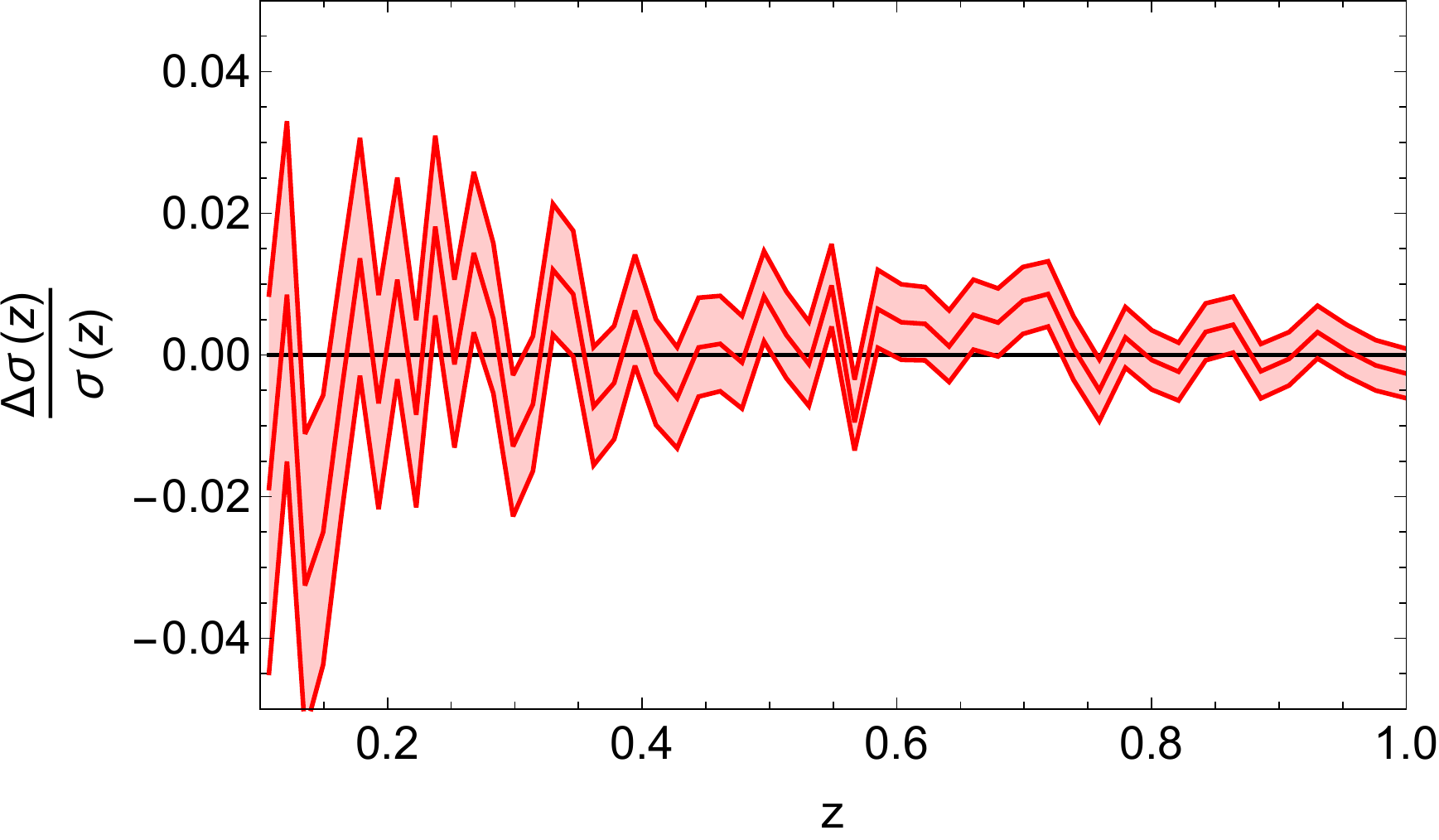}
   \caption{Precision on the estimate of the density variance, $(\hat\sigma_{\rm ML}(z)-\sigma(z))/\sigma(z)$, in a Euclid-like survey containing about 900,000 spheres of comoving radius $R=10$Mpc$/h$ regularly drawn between redshift $0.1$ and $1$.  As the number of spheres increases with accessible volume and therefore with redshift, the precision on the density variance is better at high redshift, from a few percents at $z=0.1$ to sub-percent accuracy at $z=1$.
   \label{fig:MLsigma-Euclidlike}}
\end{figure}

Let us now conduct the following fiducial experiment. In order to mimic a Euclid-like survey, 
let us consider redshifts between 0.1 and 1 binned so that the comoving distance of one bin is $40$Mpc$/h$, and  draw regularly spheres of radius $R=10$Mpc$/h$ separated by $d=40$Mpc$/h$ (hence we  ignore neighbouring spheres  and assume that the spatial correlations are negligible).
For a 15,000 square degree survey, it yields 50 bins of redshift ($z_{i}$) with a number of spheres ranging from about $N_{1}=800$ (at $z_{1}=0.1$) to $N_{50}=45,000$ (at $z_{50}=1$) for a total of almost 900,000 supposingly independent spheres.
In this experiment, we assume that the model for the density PDF is exact and that the variance (which is a free parameter) is related to the growth rate  by linear theory.
At each redshift, we can reconstruct the variance by measuring the full PDF
 \begin{equation}
\hat \sigma_{\rm ML}(z_{j})= \arg \max_{\sigma} \left\{\prod_{i=1 }^{N_{j}}   \mP\left(\rho_{i,j}|\sigma
 \right) \right\} \,.
 \end{equation} 
As an illustration, the reconstructed $\hat\sigma_{\rm ML}$ is shown in Figure~\ref{fig:MLsigma-Euclidlike}. A typical  precision on $\sigma$ of a percent is found. Note that, as expected, the reconstruction is more accurate at higher redshift where the accessible volume and therefore the number of spheres is larger.

\begin{figure}
\includegraphics[width=\columnwidth]{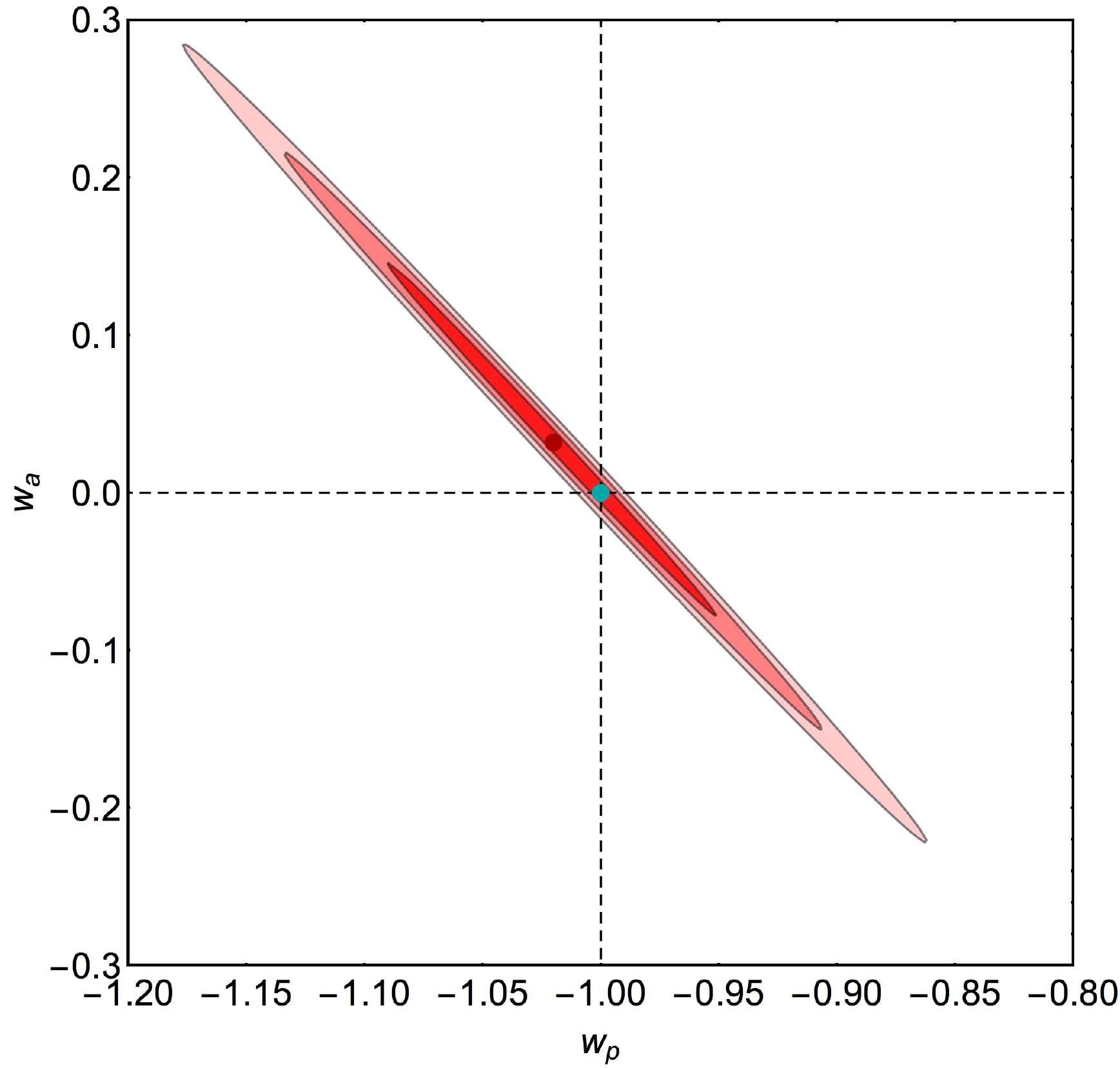}
   \caption{Contraints on the equation of state of dark energy in a Euclid-like survey containing about 900,000 spheres of comoving radius $10$Mpc$/h$ regularly drawn between redshift $0.1$ and $1$.  Contours at one, two and three sigmas are displayed with shaded areas from dark to light red. The dark red and cyan dots correspond to the recovered most likely solution and the input value respectively. The resulting precision on $w_{p}$ and $1+w_{a}$ is respectively at about the
   $5\%$ and $10\%$ level.
   \label{fig:MLeos-Euclidlike}}
\end{figure}

In order to get constraints on the equation of state of dark energy, we compute the log-likelihood of the $\sim 900,000$ measured densities $\{\rho_{i,j}\}_{{1\leq i\leq N_{j}},{1\leq j\leq 50}}$ given models for which $w_{p}$ and $w_{a}$ vary
\begin{equation}
{\cal L}(\{\rho_{i,j}\}|w_{p},w_{a})=\sum_{j=1}^{50}\sum_{i=1}^{N_{j}}\log {\cal P}(\rho_{i,j}|z_{j},w_{p},w_{a})\,,
\end{equation}
where ${\cal P}(\rho|z,w_{p},w_{a})$ is the theoretical density PDF at redshift $z$ for a cosmological model with dark energy e.o.s parametrized by $w_{p}$ and $w_{a}$.
Optimizing the probability of observing densities $\{\rho_{i,j}\}_{{1\leq i\leq N_{j}},{1\leq j\leq 50}}$  at redshifts $\{z_{j}\}_{{1\leq j\leq 50}}$ with respect to  $(w_{p},w_a)$,  
yields a maximum likelihood estimate for the dark energy equation of state parameters
 \begin{equation}
 (\hat w_p,\hat w_a)= \arg \max_{w_p,w_a} \left\{{\cal L}(\{\rho_{i,j}\}|w_{p},w_{a}) \right\} \,. \label{eq:defDop2t}
 \end{equation}
The resulting $\alpha=1,2,3$ sigma contours are shown in Figure~\ref{fig:MLeos-Euclidlike} and correspond to the models for which   {\red ${\cal L}(\{\rho_{i,j}\}|w_{p},w_{a})=\max_{w_{p},w_{a}} {\cal L}(\{\rho_{i,j}\}|w_{p},w_{a}) +\log \left(1-\rm{Erf}(\alpha/\sqrt 2)\right)$.}
Modulo our assumptions, this maximum likelihood method  allows for constraints on $w_{p}$ and $1+w_{a}$ at a few percent
level making it a competitive tool for the analysis of future Euclid-like surveys. In practice, it is expected that various uncertainties will degrade the accuracy of the proposed method (uncertainties on the model itself at low-redshift\footnote{\red For instance, it is clear that linear prediction for the density variance is not sufficient on 10 Mpc$/h$ scale at $z=0$ but note that the number of spheres used here is much bigger at high redshift where linear theory is more accurate so that the error made at low redshift should only have a small effect in our configuration. However, if needed, it should be straightforward in this formalism to add loop corrections to the variance, following for instance \cite{2010JCAP...02..021J} which uses second order perturbation theory to predict the non-linear evolution of the amplitude of density fluctuations .}, galaxy biasing, redshift space distortions, observational biases, etc). Accounting for these effects requires further works, beyond the scope of this paper. 

Our main conclusion should still hold once all those effects are accounted for, namely that there is a  significant gain in using the full knowledge of the PDF --  therefore relying on a maximum likelihood analysis -- when contrasted  to the direct measurements of cumulants (variance but also skewness, kurtosis, etc). This method is in particular less sensitive to rare events which can significantly biased the measurements of cumulants.  Large-deviation theory will therefore allow us to get tighter cosmological constraints. In particular, it has to be noted that in this work, only the redshift-dependent density variance varies but in principle we could also allow for variations of the linear power spectrum and dynamics of the spherical collapse and therefore modified gravity scenarios.

\section{Conclusion}
\label{sec:conclusion}

Recently, \cite{Uhlemann16}
presented an analytic expression for the  PDF of the density within  concentric spheres for a given cosmology as a function of redshift in the mildly non-linear regime.
Relying on such a expression, an illustrative very simple  fiducial dark energy experiment was 
  implemented while fitting, in the maximum likelihood sense,  a set of measured densities covering a range of 
  redshift, and  seeking the  best dark energy model consistent with such a set. 
It was  demonstrated through this experiment that the maximum likelihood estimator based on the  analytic  PDF for the density within
 spherical cells yields  a more accurate un-biased estimate for the redshift  evolution of the cosmic variance
than the usual sample variance, as expected for non-Gaussian statistics.
The optimality of the maximum likelihood estimator was qualified perturbatively in Appendix~\ref{sec:comparing}. 

While it is clear that the above-described  experiment is overly simplistic in many respects,
it serves as a mean to demonstrate the predictive power  of equations~(\ref{PDFfromPsi2})-(\ref{eq:defsigma}).
  They  capture  the essence of  why these PDFs outperform naive estimates based on Gaussian statistics,  and motivates the publication of the corresponding 
simple  {\tt Mathematica} code which is provided online as described in Appendix~\ref{sec:code}. 

In order to provide a more realistic framework for such a dark energy experiment,  one would need to also account for galactic bias and redshift space distortions\footnote{Thanks to the generality of the formalism for density in spheres it can be also applied to any function of the dark matter density including a local bias expansion to model the galaxy density field. For existing studies on the effect of galactic bias on the PDF see \cite{Kauffmann97,Weinberg04,Kang07,Fry11} together with \cite{Scherrer01} for the influence of redshift space distortions.}, sampling error and 
 masking. One would then need to define the optimal sampling strategy while varying the sphere's radii and their redshift evolution given the geometry of the survey, and the rate of acceleration of the Universe\footnote{\red Finding the optimal estimator to choose among all possible radii and number of spheres is still an open question. This situation is quite similar to the issue of optimal smoothing and correlations in the so-called $P(D)$ analysis (see for instance \cite{2014MNRAS.440.2791V} and references therein) which allows an estimate of source count below the confusion limit.}. This probe should of course be coupled to other existing probes whose figure of merit is not degenerate with respect to figure~\ref{fig:MLeos-Euclidlike}, so as to tighten the constraints on the equation of state of dark energy.
On the theoretical side, an obvious extension of the present work would be to consider the impact of considering only subsets of the fields, e.g.
under-dense regions \citep[e.g.][and references therein]{Bernardeau15}  while making use of the joint PDF for the density within multiple concentric spheres. Applying the  Large Deviation theory to 2D cosmic shear maps should also allow us to model 
the statistics of projected density profiles, which could be used for weak lensing studies.

Up to now, we assumed for simplicity that the different densities in spheres were uncorrelated. An improvement would be to  correct for the correlation between not-so-distant spheres,
as quantified in \cite{codis16} using bias functions which can also be predicted in the large-deviation regime. 
If we consider a  configuration made of $\rm N$  spherical cells  which centres are separated by 
distances $r_{\rm{IJ}}$, \cite{codis16} showed that
the joint PDF $\mP_R(\rho_{1},\dots,\rho_{\rm{N}};r_{\rm{IJ}}) $ of the density $\{\rho_{\rm I}\}_{{\rm I<N}}$ in the large-separation limit,
where $r_{\rm{IJ}}\gg R$, can be written as
\begin{equation}
\mP(\rho_1,\dots,\rho_{\rm N};
\{
r_{\rm{IJ}}
\}) 
\!=\!\nonumber
\prod_{\rm I=1}^{\rm{N}} \mP(\rho_{\rm I})
\!\left[\!1+\!
\sum_{\rm{I<J}}
b(\rho_{{\rm I}}) b(\rho_{{\rm J}}) \xi(r_{\rm{IJ}})
\!\right]\!, \label{eq:fullPDFlargeseparation}
\end{equation}
where $\Pi_{\rm I} \mP(\rho_{\rm I})$ is the product of one-point PDFs,
 $\xi(r)$ is the underlying dark matter correlation function, $r_{\rm{IJ}}$ is the separation between the cells $\rm I$ and $\rm J$ of radius $R$ and $b(\rho)$ is a density bias function 
 given by
  \begin{equation}
b(\rho)= \frac{\zeta^{-1}(\rho)}{\sigma^2(R\rho^{1/3})} 
\,,
 \end{equation}
  for low densities  \citep[see][for a more general expression]{codis16}.
 Taking into account the spatial correlations described in equation~(\ref{eq:fullPDFlargeseparation}) is left for future work.

\vspace{0.5cm}

{\bf Acknowledgements:}  
 This work is partially supported by the grants ANR-12-BS05-0002 and  ANR-13-BS05-0005 of the French {\sl Agence Nationale de la Recherche}.
 CU is supported by the Delta-ITP consortium, a program of the Netherlands organization for scientific research (NWO)  funded by the Dutch Ministry of Education, Culture and Science (OCW).
 SC and CP thank the CFHT and Susana for hospitality and Karim Benabed for insightful conversations. 
SC and CU thank Martin Feix and the participants of the workshop ``Statistics of Extrema in Large Scale Structure'' for fruitful discussions when this work was completed.
SC also thanks the University of British Columbia for hospitality and Douglas Scott for interesting discussions.
\vskip -0.75cm
\bibliographystyle{mn2e}
\vskip -0.75cm

\bibliography{LSStructure}

\appendix
\section{LSSFast Package}
\label{sec:code}

\begin{figure*}
\includegraphics[width=\columnwidth]{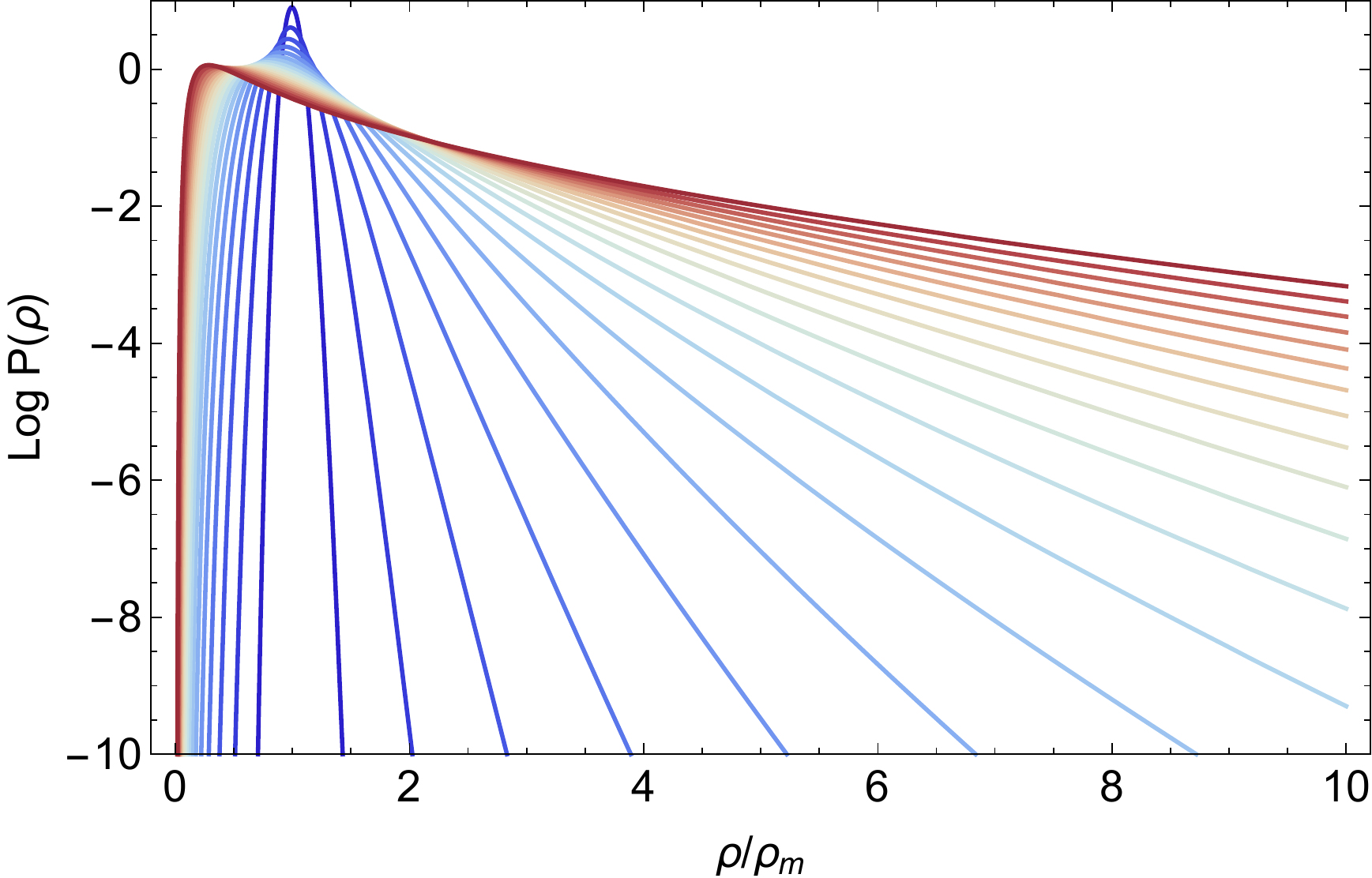}
\includegraphics[width=\columnwidth]{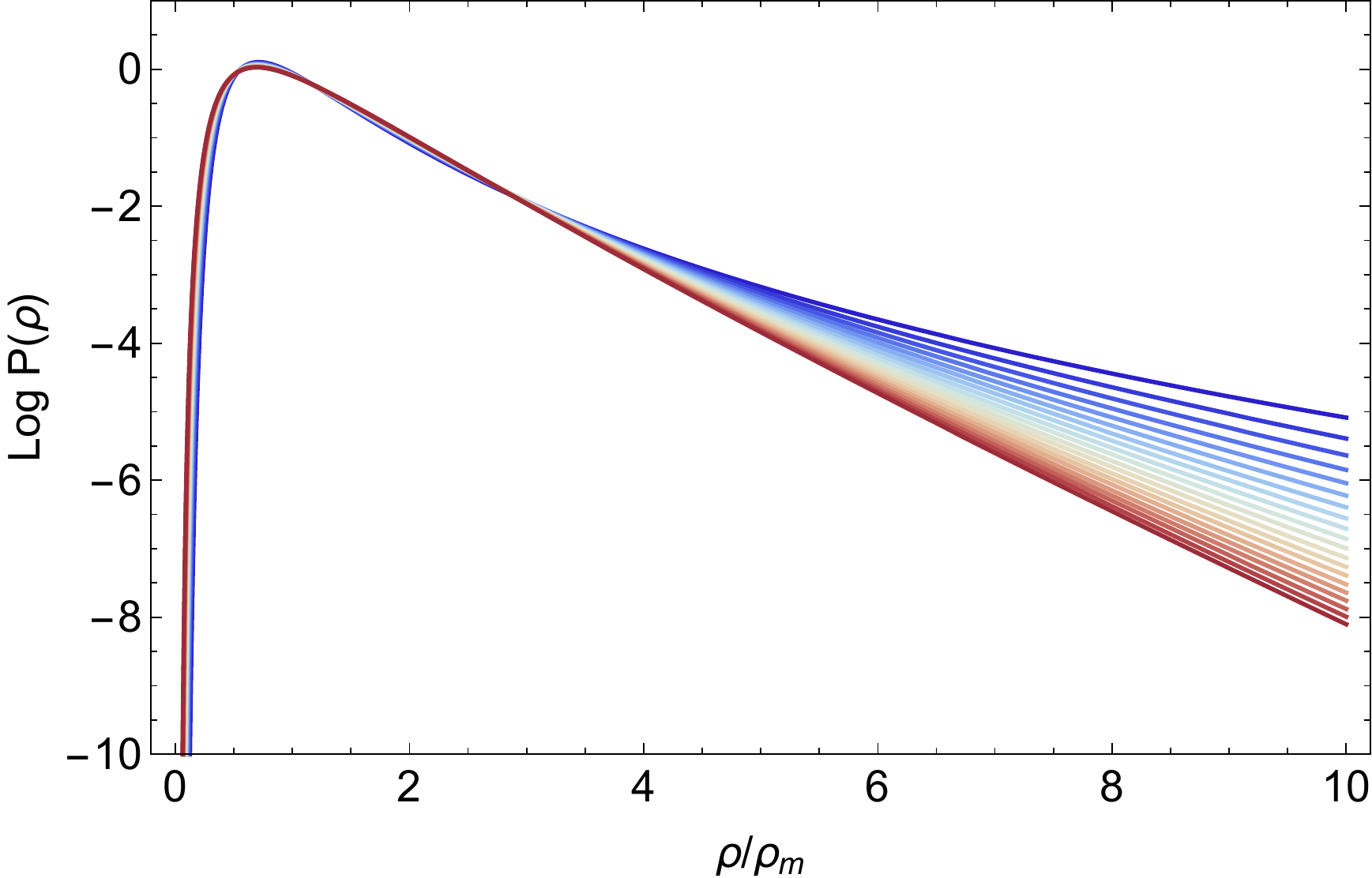}
   \caption{{\it Left-hand panel:} PDF of the density in cells of radius 10Mpc$/h$ as a function of $\sigma$ from 0.05 (dark blue) to 1 (dark red) computed from the {\tt LSSFast} code. In the low $\sigma$ limit,
   the PDF is essentially Gaussian, while it becomes very skewed at larger $\sigma$ (in red). Here the linear power spectrum is for a WMAP7 cosmology and the spherical collapse dynamics is parametrized by $\nu=21/13$. {\it Right-hand panel: }same as left-hand panel but for $\sigma=0.5$ and cells of radius from R=1Mpc$/h$ (blue) to 20 Mpc$/h$ (red). 
   \label{fig:buddle}}
\end{figure*}

The density PDF for power-law and arbitrary power spectra are made available in the {\tt LSSFast} package distributed freely at \url{http://cita.utoronto.ca/~codis/LSSFast.html}.
Two versions of the code are presented. The simpler version, {\tt $\rho$PDFns}, assumes a running index, meaning that the variance is given by
\begin{equation}
\sigma^{2}(R)=\frac{2\sigma^{2}(R_{p})}{\left( R /{R_{p}}\right)^{3+n_{s}+\alpha}+\left( R/ {R_{p}}\right)^{3+n_{s}-\alpha}}\,,
\end{equation} 
where $\alpha$ can be non-zero to take into account the variation of the spectral index $n_{s}$. The density PDF is analytically computed from equation~(\ref{PDFfromPsi2}) and numerically regularised to enforce the normalisation, the mean and the variance. This code is very efficient and runs in about one second on one processor for one evaluation. Note that the function {\tt $\rho$PDFns} takes three arguments, $\rho$, $\sigma$ and $n_{s}$, and has one option, $\alpha$.

The second version of the code, {\tt $\rho$PDF}, uses input from  {\tt camb} \citep{camb} and can therefore be applied  to arbitrary power spectra. In this case, the function $\sigma^{2}(R)$ is tabulated using equation~(\ref{eq:defsigma}). Once this tabulation is done (typically one minute on one processor), each evaluation of the PDF takes about the same time as for the power-law case ($\approx 1$ sec).

As an illustration, Figure~\ref{fig:buddle} presents 
a bundle of density PDFs computed with {\tt LSSFast} for different variances and radii and a $\Lambda$CDM power spectrum. At low $\sigma$ or equivalently high redshift (in blue), the PDF is almost Gaussian and concentrated around the mean but as the variance grows (i.e at smaller redshift), it becomes increasingly skewed and broad, while the maximum of this PDF 
is shifted in under-dense regions as voids are getting emptier and nodes denser. 

\section{Comparing estimators}
\label{sec:comparing}

Let us compare, from a theoretical point of view, our two different estimators of the variance of the density field.
The first one is the sample variance defined as usual as
\begin{equation}
\label{eq:sigA}
\hat \sigma_{A}^{2}=\frac 1 N \sum_{i=1}^{\rm N}(\rho_{i}-1)^{2}\,,
\end{equation}
where the measured densities $\rho_{i}$ in $\rm N$ cells are rescaled by their mean.
It can be shown  that the expectation value of this estimator is the underlying ``true'' variance $\left\langle\hat\sigma^{2}_{A}\right\rangle=\sigma^{2}$. The typical variance of this estimator can also be computed as
\begin{equation}
\left\langle(\hat\sigma^{2}_{A}-\sigma^{2})^{2}\right\rangle=\frac{S_{4}\sigma^{6}+2\sigma^{4}}{\rm N}\,,
\end{equation}
where $S_{4}$ is the kurtosis $\left\langle\rho^{4}\right\rangle_{c}/\sigma^{6}$.

Let us now compare the estimator of the sample variance to the maximum likelihood estimator 
\begin{equation}
\label{eq:sigML}
\hat\sigma_{\rm ML}^{2}=\argmax_{\tilde\sigma^{2}}\prod_{i=1}^{\rm N}\mP(\rho_i |\tilde\sigma^{2})\,,
\end{equation}
where the modelled PDF $\mP(\rho |\tilde\sigma)$ 
depends on the value of the variance $\tilde \sigma$.
It is a well-known result \citep{stuart2009kendall} that the estimate of the variance in this case converges towards the true value $\sigma$ in a probabilistic sense (via the so-called relation of consistence).
In addition, there is an asymptotic normality in the sense that the asymptotic distribution of $\sqrt {\rm N} (\hat\sigma_{\rm ML}^{2}-\sigma^{2})$ is a Gaussian of zero mean and variance given by the inverse Fisher information
\begin{equation}
\Sigma_{\rm ML}^{2}=-1/\left\langle{\cal L}''(\rho|\sigma^{2}) \right\rangle\,,
\end{equation}
where ${\cal L}''(\rho|\sigma^{2})=\partial^{2}\log {\cal P}(\rho|\sigma^{2})/\left(\partial \sigma^{2}\right)^{2}$.
Assuming the PDF is given by  equation~(\ref{PDFfromPsi2}),
the Fisher information of $\rho$ can be easily computed
\begin{equation}
\label{eq:Fischer-PT}
-\left\langle{\cal L}''(\rho|\sigma^{2}) \right\rangle=
\frac{2\left\langle \Psi\right\rangle}{\sigma^{6}}-\frac 1 {2\sigma^{4}}
\,.
\end{equation}
Note that here we did not take into account the normalisation of the PDF but as its $\sigma$ dependence is rather small, its contribution is expected to be negligible.
The mean of the rate function appearing in equation~(\ref{eq:Fischer-PT}) can then be perturbatively computed
\begin{align}
\left\langle \Psi(\rho)\right\rangle &=\left\langle [\zeta^{-1}(\rho)]^2\frac{\sigma^{2}(R)}{\sigma^{2}(R\rho^{1/3})}\right\rangle\nonumber\\
&=\frac 12\sigma^{2}+\frac1 {24}\sigma^{4} (5 S_3^2-3 S_{4})+
{\cal O}(\sigma^{6})\,.
\end{align}
Using this perturbative approach, one can show that the inverse Fisher information is given by
\begin{equation}
\label{eq:PTsML}
\Sigma_{\rm ML}^{2}\!=\!\frac{-1}{\left\langle{\cal L}''(\rho|\sigma^{2}) \right\rangle}\!=
\!2\sigma^{4}+\sigma^{6}\left(S_{4}-\frac{5S_{3}^{2}}{3}\right)+{\cal O}(\sigma^{8})
\,.
\end{equation}
As expected, the variance obtained for the sample variance estimator is larger than the one obtained in the maximum likelihood approach by a factor 
$-{5S_{3}^{2}}\sigma^{6}/3$ proportional to the  square of the skewness  of the density field.  In particular, we recover that both approaches are equivalent only in the Gaussian limit. As soon as non-Gaussianities appear, the sample variance is not optimal anymore.

\begin{figure}
\includegraphics[width=\columnwidth]{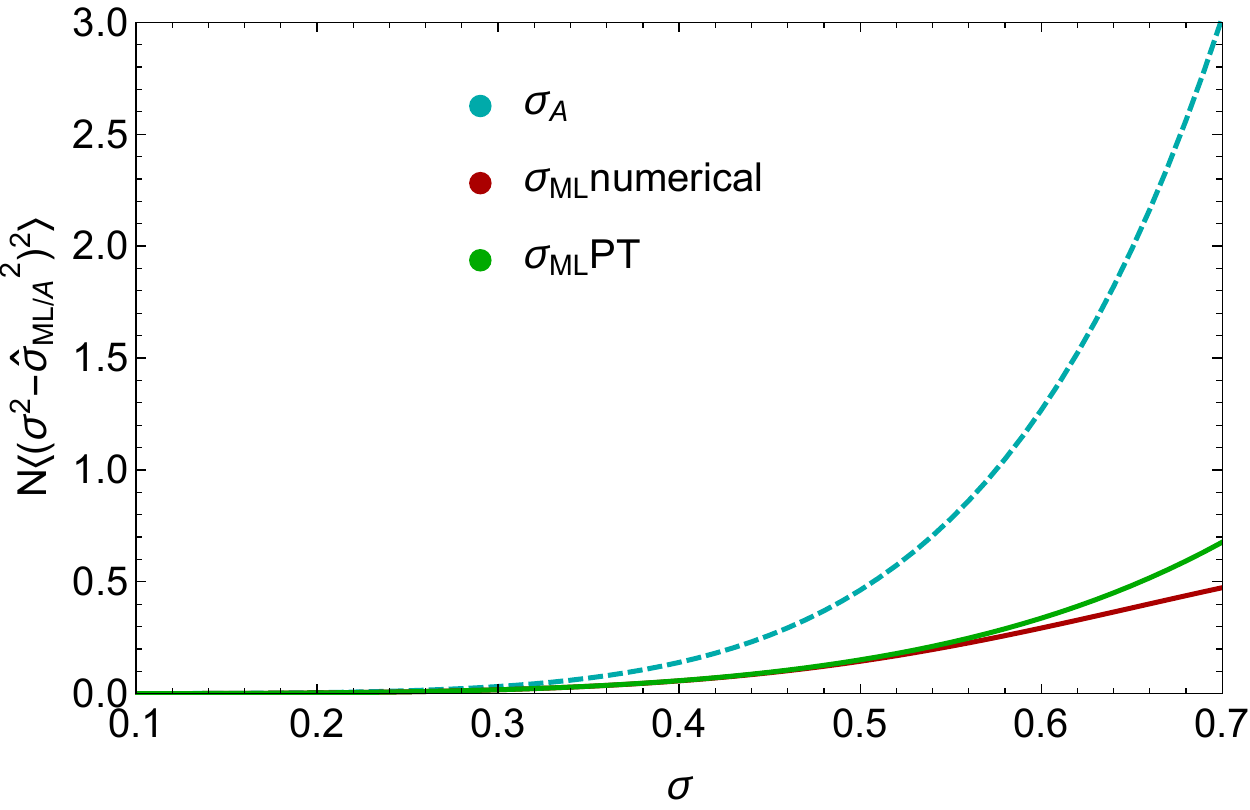}
   \caption{Variance of the density variance estimator for the arithmetic estimator ($\hat{\sigma}_{\rm A}$) in cyan compared to the variance of the Maximum likelihood estimator ($\hat{\sigma}_{\rm ML}$) in red estimated numerically in the asymptotic limit (N goes to infinity) and estimated perturbatively by equation~(\ref{eq:PTsML}) in green. As expected, both estimators are equivalent at small $\sigma$ when the PDF is nearly Gaussian but as non-linearities arise, the performance of the arithmetic variance decreases as it is not optimal anymore. 
   \label{fig:MLvsA}}
\end{figure}

Figure~\ref{fig:MLvsA} compares, in the asymptotic limit (N going to infinity), the sample variance to the maximum likelihood variance obtained by a non-perturbative approach where $\left\langle{\cal L}''(\rho|\sigma^{2}) \right\rangle$ is integrated numerically. It shows that the sample variance is sub-optimal when $\sigma$ increases while the PDF becomes non-Gaussian. This result is in good agreement with the Monte-Carlo estimate shown in Figure~\ref{fig:fit-error}. The perturbative analytical prediction given by equation~(\ref{eq:PTsML}) is found to reproduce well the expectation at low $\sigma$ ($\lesssim 0.5$).

Note that, in order to account for the spatial correlations between the cells (measured densities are not independent), 
one can model the joint statistics in the large-separation limit.
In that case, an additional error is made so that for instance  the sample variance is given by
\begin{equation*}
\left\langle(\hat\sigma^{2}_{A}-\sigma^{2})^{2}\right\rangle=
\frac{S_{4}\sigma^{6}+2\sigma^{4}}{\rm N}+
\frac{\sum_{\rm{I\neq J}}\xi(r_{\rm{IJ}})}{\rm N^{2}}\left(B_{2}^{2}-4B_{2}+4\right)\,
\end{equation*}
where $B_{2}=\left\langle b(\rho) \rho^{2}\right\rangle$, $\left\langle b(\rho) \rho\right\rangle=1$, $\left\langle b(\rho)\right\rangle=0$ and $b(\rho)$ is the density bias defined in \cite{codis16}.

\end{document}